# Specification Test Compaction for Analog Circuits and MEMS*

Sounil Biswas[†], Peng Li[‡], R. D. (Shawn) Blanton[†] and Larry T. Pileggi[†]

[†]Dept. of Electrical and Computer Eng.
Carnegie Mellon University
Pittsburgh, PA 15213
{sbiswas,blanton,pileggi}@ece.cmu.edu

[‡]Dept. of Electrical Eng.
Texas A&M University
College Station, Texas 77843
pli@ee.tamu.edu

## Abstract

*Testing a non-digital integrated system against all of its specifications can be quite expensive due to the elaborate test application and measurement setup required. We propose to eliminate redundant tests by employing $\varepsilon$-SVM based statistical learning. Application of the proposed methodology to an operational amplifier and a MEMS accelerometer reveal that redundant tests can be statistically identified from a complete set of specification-based tests with negligible error. Specifically, after eliminating five of eleven specification-based tests for an operational amplifier, the defect escape and yield loss is small at 0.6% and 0.9%, respectively. For the accelerometer, defect escape of 0.2% and yield loss of 0.1% occurs when the hot and cold tests are eliminated. For the accelerometer, this level of Compaction would reduce test cost by more than half.*

## 1. Introduction

In contrast to digital circuit test, there are no universally accepted fault models for non-digital components. Consequently, validating the circuit specifications continues to be the most accepted form of non-digital component test [1]. However, the number of circuit specifications to be verified has increased significantly over the past few decades. In addition, the cost to verify each specification can be exorbitant due to elaborate test setup, test pattern application and measurement. In order to reduce test cost, only a subset of the specifications are actually verified in practice [2]. The subset of specifications used for testing is selected based on the experience of designers and test engineers. Such an *ad hoc* selection of the specification set can lead however to an unpredictable amount of yield loss and defect escape.

A desired solution to this problem is to develop a cost-effective test methodology that can verify all the circuit specifications. Several efforts have been published that address this goal. For example, various low-cost, built-in self-test (BIST) solutions have been proposed that reduce test cost by eliminating external test equipment [3, 4]. Alternately, efforts have been made to use digital circuitry to generate tests and analyze test data for mixed-signal circuits [5]. Work in [6] describes a test sequence generation approach that can implicitly verify all the circuit specifications. In this approach, test cost is reduced by using tests that are not derived from the specifications.

We propose to achieve the goal of obtaining a cost effective test set that can verify all the device specifications by identifying and eliminating the redundancy in the complete specification test set using statistical learning. Beginning the test set compaction procedure from the complete specification-based test set guarantees no initial defect escape or yield loss'. Redundant tests are eliminated until the prediction error exceeds a user-defined tolerance. During test set compaction, a statistical model is derived to predict pass/fail behavior based on the remaining tests. Since the amount of prediction error of the derived statistical model is constrained, the procedure enables controlled yield loss and defect escape in contrast to *ad hoc* test compaction. In addition, when the statistical model makes an inconsistent prediction, the tested part is deemed to belong to a "guard-band region". Devices in the guard-band region can be further tested to prevent unnecessary yield loss.

The rest of the paper is organized as follows: In Section 2, we introduce the necessary background for the problem addressed. Section 3 contains a description of the statistical learning based specification test set compaction procedure. Section 4 describes solutions to the various roadblocks encountered during test compaction. In Section 5, we describe simulation results showing the viability of the test set compaction procedure for an operational amplifier and a MEMS accelerometer. Finally, in Section 6, our work is summarized and future work is described.

## 2. Preliminaries

In this section, we present the terminology and background necessary for discussion of the specification-based test compaction methodology.

### 2.1. Definitions

A specification $s_i$ for a device under test (DUT) is a performance parameter of the circuit that must be measured and verified. A specification test $t_i$ for $s_i$ includes setup for application of stimuli, application of the stimuli and analysis of the responses to obtain the value of $s_i$ for a particular manufactured device. We define a device instance to be **good** if all the specification values $S = \{s_1, s_2, \ldots, s_n\}$ obtained from the specification-based tests $T = \{t_1, t_2, \ldots, t_n\}$ fall within a set of desired ranges $R = \{r_1, r_2, \ldots, r_n\}$, where $r_i$ is the acceptable range for specification $s_i$. A device instance is deemed **bad** or **faulty** if any of the values $s_i$ obtained from the specification tests T fall outside the corresponding

---

*This research is supported by the MARCO/DARPA C2S2 and GSRC programs under contracts 2003-CT-888 and 2003-DT-660, respectively.

'There can, of course, be yield loss and defect escape based on the type of test used for each specification however.

range $r_i$. The process of finding if a device is good or bad is called the *pass/fail analysis.*

In this paper, we achieve test compaction by identifying and eliminating tests $T_{red} = \{t_p, \ldots, t_q\}$ for redundant specifications $S_{red} = \{s_p, \ldots, s_q\}$ from the complete set of specification-based tests T. It is important to note however that a greater level of compaction may be possible if the compaction process is performed at the test level as opposed to the specification level.

### 2.2. Statistical Learning and SVM

The following terminology is adapted from [7,8]. Given a set of data points $\{(X_1, y_1), \ldots, (X_l, y_l)\}$, where $X_k = [X_k^1, \ldots, X_k^m]$ is an m-dimensional input and $y_k$ is a one-dimensional output, Statistical (Machine) Learning is the process of building a model for the unknown relation $y = f(X)$ using the given set of data points. The data that is used to build the model is called the *training data,* while the data used to verify the model is called the *test data.*

In this work, we have used $\varepsilon$-Support Vector Machines (SVM) [7, 8] for statistical learning based model derivation. The goal in $\varepsilon$-SVM is to build a hyper-plane that divides the input $X_k$'s into two classes that pertain to $y_k = 1$ and $y_k = -1$. Given a set of training data $\{(X_1, y_1), \ldots, (X_l, y_l)\}$ and a predefined error limit $\varepsilon$, this classification is achieved by building a function $f(X)$ to estimate $y_k$ such that: (1) for each $(X_k, y_k)$, $f(X_k)$ has at most $(\varepsilon + \xi_k)$ deviation from $y_k$, where $\xi_k \geq 0$ is an unbound error, (2) $\sum_{k=1}^{l} \xi_k$ is minimized, and (3) the number of local maxima and minima in $f(X)$ is minimized. This means that the error in estimating $y_k$ at most data points $(X_k, y_k)$ in $\{(X_1, y_1), \ldots, (X_l, y_l)\}$ is controlled to be less than $\varepsilon$. However, there can be $(X_k, y_k)$'s where $|y_k - f(X_k)| > \varepsilon$, and the additional error is denoted as $\xi_k$. The value $e_m = y_k - f(X_k)$ is the *model error* for $f(X)$ at $(X_k, y_k)$.

Since testing is intrinsically a pass/fail classification problem, $\varepsilon$-SVM is especially suited for the test compaction methodology. Additionally, $\varepsilon$-SVM gives sufficiently high accuracy in reasonable time making its repeated application feasible. Finally, $\varepsilon$-SVM can efficiently handle the huge data set that is required during test compaction (details are given in Section **4.1).**

## 3. Test Set Compaction Methodology

To remedy the drawback of traditional *ad hoc* elimination of tests, we propose a systematic approach to prune the test set using $\varepsilon$-SVM.

### 3.1. Training Data Generation

Training data is generated using Monte-Carlo simulations of devices with random variations imposed on various device parameters. This approach most likely creates some devices that are not possible within a real manufacturing environment. However, a test compaction methodology that works well here will also perform well for realistic devices. In addition, test compaction using training data generated from device simulation is significantly less expensive as compared to actual tests applied to a manufactured chip [9]. However, the downside is that simulation data is likely less accurate.

Figure **1** shows a block diagram illustrating the process of training data generation for $N$ device (training) instances. In the process of training data generation, the device description, the list of device specifications that are to be considered for compaction, and a manufacturing process model for the device serves as input. The device representation is altered based on the process model. Each altered device representation is known as a *training instance.* Based on the set of specifications under consideration, the device instance is set up, simulated, and measurements are derived from simulation results. This measured data is stored as training data. The whole process is repeated until the number of training instances simulated equals $N$, the number of training instances desired.

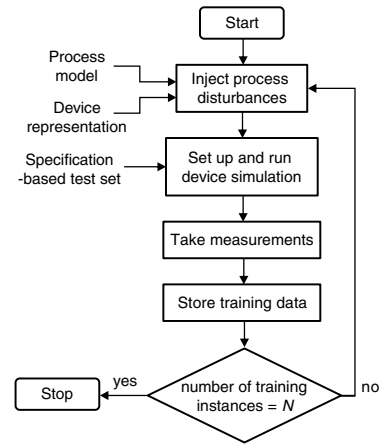

**Figure 1.** A flowchart description of training data generation.

### 3.2. Test Set Pruning

The goal of a systematic test set compaction approach is to offer control over yield loss and defect escape while test cost is reduced. Figure **2** illustrates how the proposed test compaction methodology achieves this goal. The process of test compaction begins with the complete specification-based test set, the training data set and acceptability ranges for each specification. Based on the range data, the pass/fail nature of each device instance is obtained and added to the training data. Next, the test compaction process starts with the complete test set $T$, meaning that $T_{red}$ is empty which implies no initial yield loss or defect escape. A test $t, \in \{T - T_{red}\}$ is then selected for potential elimination and is removed from the training data set. Using $\varepsilon$-SVM, the reduced training data is utilized to build a model for predicting pass/fail for the set of eliminated specifications $S_{red}$. The total number of device instances that are misclassified with respect to $S_{red}$ defines the *prediction error $e_p$.* If it is possible to build a $\varepsilon$-SVM model to have a prediction error on test data to be lower than a *user-defined tolerance* $e_T$, the chosen test is deemed redundant and is permanently deleted from the set of specification-based tests. If, on the other hand, the model cannot be created with sufficient accuracy, the test is deemed necessary and is added back to the set of specifications to be explicitly tested. This process is continued until each test $t_r \in T$ is examined. At the end of this process, a compacted set of specification-based tests and a model predicting the pass/fail behavior of the device based on the compacted set of tests is obtained.

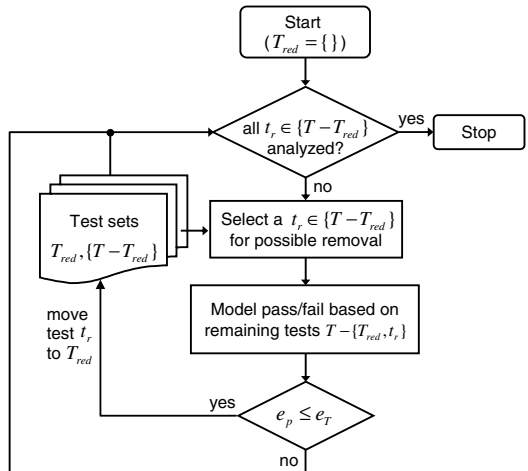

**Figure 2.** A flowchart description of the process of specification test compaction.

The aforementioned procedure described is obviously greedy which means the optimality of the solution depends on the order in which the tests are examined. An alternative approach involves assessing the number of training instances successfully classified by each specification to decide test order. Yet another approach focuses on clustering specifications based on an estimate of their mutual dependence. The resulting clusters would be used to decide test order. In our case, we analyze device functionality to decide the order of the tests.

### 3.3. Application on the Tester

A compacted test set is the set of specification-based tests that must be applied to the DUT. However, the acceptability ranges are no longer valid. In fact, new boundary surfaces for the acceptability ranges of the compacted test set are created by the statistical model.

Consider three specifications $s_1$, $s_2$ and $s_3$, where $s_3$ can be expressed as $s_3 = f(s_1, s_2)$. After elimination of the test for redundant specification $s_3$, new acceptability ranges $r_1$ and $r_2$ for specifications $s_1$ and $s_2$ describing the pass/fail behavior of the DUT are created. Essentially, the acceptability ranges change because there can be device instances that fall within $r_1$ and $r_2$ but not within $r_3$, meaning the instance is faulty. The change in $r_1$ and $r_2$ is illustrated in Figure 3. The rectangular box in Figure 3 represents the acceptability ranges $r_1$ and $r_2$, while the shaded region is the derived area defining good devices when the test $t_3$ for $s_3$ is eliminated. It is therefore necessary to provide the tester with this additional information for correct pass/fail analysis of the manufactured devices. One solution is to provide the tester with the statistical model derived from the training data. However, this may require a significant amount of additional tester resources. Alternately, we **propose** to divide the space of compacted specifications into grids and assign a good or bad attribute to each grid cell based on information gathered from the statistical model. This creates a lookup table that can be added to the tester program with little additional cost.

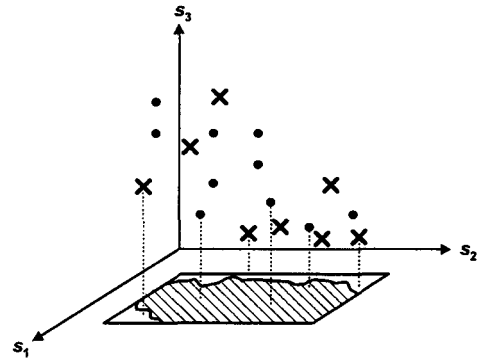

**Figure 3.** An illustration of the derivation of new acceptability ranges for pass/fail analysis. The dots represent "good" device instances while the crosses represent "bad" device instances.

## 4. Statistical Learning for Test

Given the overview of the proposed test compaction methodology in Section 3, here we elaborate upon the issues associated with applying statistical modeling to the manufacturing test of analog and MEMS DUTs.

### 4.1. Classification versus Regression

The traditional approach for statistical learning based test has focus on regression [6,10]. However, regression is a multi-dimensional problem, where each dimension is a specification. Consequently, the training data required to adequately cover the multidimensional space for building the regression model can be extremely large. We observe that the test problem is a simple pass/fail classification issue. Additionally, classification only requires adequate coverage of the class boundaries, thus significantly reducing the training data required for modeling. Therefore, $\varepsilon$-SVM based classification is an appropriate choice for the test compaction problem addressed.

### 4.2. Guard-Band Region

It is important to note that pass/fail analysis for test is a two-class classification problem. This means a DUT that has measured specification values for each specification $s_i \in S$ within $r_i$ is "good" and one that has at least one measured $s_i$ value outside $r_i$ is "bad". As a result, there is discontinuity at the boundary between good and bad. Consequently, a very small model error for devices near the boundary can lead to significant misclassification. Consider a set of devices located near the boundary of the pass/fail decision. Also assume that these devices have the same values for the set of specifications $\{S-S_{red}\}$ but have unequal values for the eliminated set of specifications $S_{red}$. All of these devices will have the same prediction, say good. However, they may not all be good devices. Assuming that their location near the pass/fail decision boundary can be represented by a normal distribution (see Figure 4(a)) means there is a significant probability that bad devices can be classified as good due to prediction error.

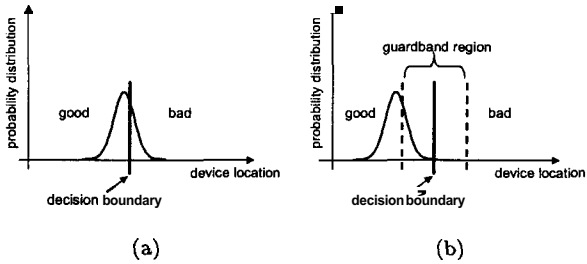

**Figure 4.** An illustration of the accuracy improvement going from (a) an unguarded boundary to (b) a guarded decision boundary. The solid line indicates the decision boundary, while the dotted lines show the guard band.

To solve this problem, we propose to guard-band the decision boundaries. Guard banding significantly reduces the probability of a good device being predicted as bad (yield loss) or a bad device being predicted as good (defect escape). Guard banding is useful because only the tail of the device distribution from a good or a bad prediction falls in the incorrect class. Consider now a set of devices near the guard-band boundary, where again all the devices have the same values for $\{S - S_{red}\}$ and different values for $S_{red}$. From Figure 4(b), we observe that only the tail of the device distribution associated with that particular prediction falls in the incorrect class. The use of a guard band therefore significantly reduces the likelihood of misprediction.

We use two classification models to build the guard-band region. The boundaries are perturbed in each direction by a preset value to create these two models. A device predicted to be good or bad by both models can be classified to be good or bad with high confidence. On the other hand, a contradiction in the predictions from the two models means that the device is in the guard-band region. The two dotted lines in Figure 4(b) are obtained by perturbing the boundary by a predefined amount and the region between the two dotted lines is the guard-band region.

Based on the number of expected devices in the guard-band region and cost, devices can be further tested to answer the pass/fail question. Since it is likely that the number of devices belonging to this class will be small, potential benefits of adaptive test can be easily obtained. Alternately, the devices in the guard-band region may be deemed as good or bad, or even classified to be of lower grade based on the quality requirements of the application.

### 4.3. Training Data Compaction and Normalization

Although more training instances is desirable for obtaining a good coverage of the input space, building a statistical model with a large data set can be quite time consuming. To alleviate this problem, we propose to compress the data set by eliminating some training instances. Specifically, the training space is divided into a grid space. Only training instances belonging to grid cells with good as well as bad training instances are retained in the new training data set. For grid cells that contain only good or bad devices, the training instances are merged by using the center point of the grid cell with the appropriate pass/fail attribute.

Because different specifications have different units and ranges of variations, it is necessary to normalize the data to ensure uniform convergence of the multi-dimensional space. Each specification $s_i$ has a predefined acceptability range $r_i$ that is used to determine if the device satisfies $s_i$. We normalize each specification value of the training instance by mapping the lower bound of the $r_i$ to zero and the upper bound to one. This means that after normalization, a good device specification value is in the range [0,1] and any value outside the [0, 1] range is bad.

## 5. Test Compaction Examples

The proposed methodology has been applied to an operational amplifier and a MEMS accelerometer. For both examples, the training data has been generated using the circuit simulator Virtuoso Spectre [11]. NODAS [12] libraries have been used to model the MEMS accelerometer. For statistical classification, we have used SVM$light$ [13] to perform $\varepsilon$-SVM.

### 5.1. Operational Amplifier

The first device considered was an operational amplifier that was not fabricated. Five thousand training instances and 1000 test instances were generated by randomly altering the MOSFET lengths and widths and capacitor values within ±10% of their nominal values. Our choice for training and test data sizes was somewhat arbitrary in that it was based on the amount of CPU time required for simulation. Recall that training instances are used for model generation, while test instances are used to verify the derived model. A guard band of ±1% of the acceptability range boundaries was defined. Guard band size was chosen to optimally trade-off the number of instances in the guard band versus prediction error. The prediction error is equated to the number of incorrect predictions made by the derived model. More specifically, yield loss is defined as the number of good devices the model predicted to be bad, and defect escape is the number of bad devices the model predicted to be good.

Table 1 lists the set of specifications considered for the operational amplifier. The specifications considered and their units are listed in Columns 1 and 2, respectively. Column 3 contains the nominal values of each specification, while Column 4 lists the range of values in which the circuit is deemed good. The corresponding yield of the operational amplifier is 75.4% and 84.8% for the training and test data, respectively.

Figure 5 plots the yield loss, defect escape and the number of devices predicted to belong to the guard-band region for increasing levels of test compaction. The x-axis shows the eliminated tests. Specifically, the plots show the various components of error as the named tests are cumulatively eliminated as we move from left to right along the x-axis. As observed from the plot in Figure **5,** a significant amount of specification compaction ($\approx 50\%$) can be obtained at a very low level of prediction error, while the number of instances predicted in the guard-band region remains stable. This means that a trade-off between the amount of test cost reduction and prediction error exists. The optimal amount of test cost reduction can be determined by setting user-defined error limits based on the application.

| Specification $s_i$ | Unit | Nominal value | Range $r_i$ |
|---|---|---|---|
| Gain | – | 14000 | 1000–20000 |
| 3-dB bandwidth | Hz | 200 | 130–10000 |
| Unity gain frequency | MHz | 2.1 | 1.7–5 |
| Slew rate | V/$\mu$s | 0.44 | 0.35–0.55 |
| Rise time | $\mu$s | 8.5 | 0.01–10.5 |
| Overshoot | – | 0.0001 | -0.00026–0.00026 |
| Settling time | ns | 895 | 1–1070 |
| Quiescent current | $\mu$A | 105 | 70–125 |
| Common mode gain | – | 0.08 | 0–0.48 |
| Power supply gain | – | 0.4 | 0–0.95 |
| Short circuit current | mA | 0.5 | 0–4.2 |

**Table 1.** The operational amplifier specifications considered for compaction, their nominal values, and corresponding acceptability ranges.

Note that the decrease in the number of devices in the guard-band that occurs for the elimination of the short-circuit test is accompanied with an increase in defect escape and therefore does not reflect an improvement in model accuracy as tests are eliminated. Of course, the unknown model error associated with prediction can cause a non-montonic change in accuracy but this is unlikely given the use of a well-chosen guard band.

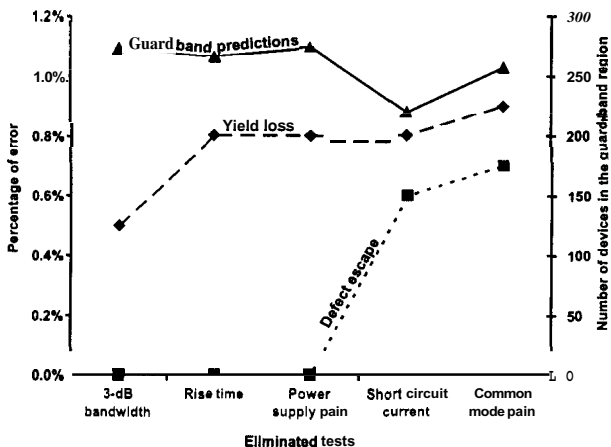

**Figure 5.** Percentages of yield loss, defect escape and number of devices predicted in the guard band as specification tests are dropped in order from left to right.

To illustrate how model accuracy is improved by the amount of training data, we have plotted trends for yield loss, defect escape, and the number of devices predicted in the guard band against number of training instances in Figure *6* when the test for 3-dB bandwidth is eliminated. The gradual reduction in yield loss and defect escape with the number of training instances shows that higher accuracy can be achieved by increasing the number of training instances.

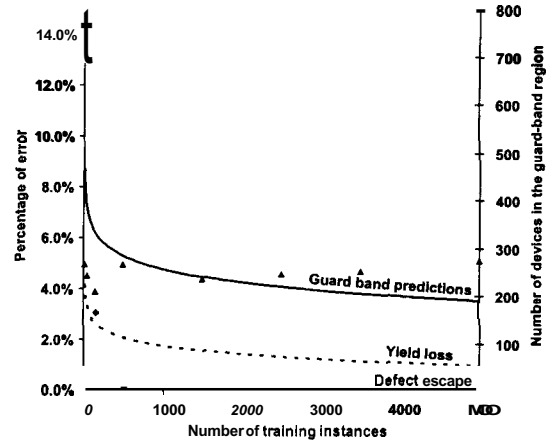

**Figure 6.** Percentages of yield loss, defect escape and number of devices predicted in the guard band with increasing number of training instances.

### 5.2. MEMS Accelerometer

MEMS accelerometers are subjected to tests at hot and cold temperatures in addition to the room-temperature tests for the same set of specifications [14]. These tri-temperature tests are quite expensive as the manufactured chip has to be heated or cooled until it reaches the desired steady-state temperature. Significant cost can therefore be avoided if temperature test outcomes can be predicted using the room-temperature test results.

We analyzed a MEMS accelerometer design from [15] using 1000 training and 1000 test instances. These instances were generated by adding $\pm 10\%$ variations to the accelerometer component lengths, widths and relative angles. Similar to the amplifier example, the number of device instances generated was limited by amount of CPU time required by Spectre to simulate an accelerometer instance. The effect of temperature is modeled as chip shrinkage or expansion, meaning the anchors of the accelerometer move towards or away from the center of the accelerometer for cold and hot temperatures, respectively. Finally, the guard-band region was defined to contain devices that have specification values within $\pm 2.5\%$ of the acceptability range boundaries. Similar to the amplifier, guard band size was chosen to optimally trade-off the number of instances in the guard band versus prediction error. Table **2** lists the accelerometer specifications that are tested at the three temperatures. The accelerometer yield is 77.4% and 79.3% for the training and test data, respectively. The hot, cold and room temperatures used for testing are 80°C, −40°C and 14.85°C, respectively.

The results of the compaction process for the hot and cold temperature tests are given in Table 3. Rows 3 through 5 in Table 3 contain respectively the percentages of the prediction error as well as devices predicted to belong to the guard-band for the elimination of cold, hot and both temperature tests.

| Specification | Unit | Nominal value | Range |
|---|---|---|---|
| Scale factor | mV/V | 9.5 | 5−30 |
| Cross-axis sensitivity | mV/V | 0.0 | -6−4 |
| Peak frequency | KHz | 5.6 | 4−6.2 |
| Quality factor | − | 2.1 | 1−2.8 |
| 3-dB bandwidth | KHz | 2.7 | 2−3.8 |

**Table 2.** Specifications for a MEMS accelerometer, their nominal values and ranges.

| Eliminated test | Defect escape (%) | Yield loss (%) | Predictions in guard band (%) |
|---|---|---|---|
| -40 | 0.1 | 0.0 | 2.6 |
| 80 | 0.1 | 0.1 | 5.8 |
| Both | 0.2 | 0.1 | 8.4 |

**Table 3.** Results for eliminating the hot and cold temperature tests for a MEMS accelerometer.

From Table 3, we observe that the hot and cold temperature tests can successfully be eliminated with very little error. Obviously this is the outcome of intrinsic functional dependence of the hot and cold temperature test outcomes on the room temperature test results. This dependence is uncovered by statistical learning. Additionally, the number of devices predicted to belong to the guard band region is small. This means that these devices can be subjected to additional test without significantly altering the reduction in test cost achieved by the proposed test compaction.

Let us examine the value of the guard band region further for the MEMS accelerometer. For the 1000 test instances considered, instances within ±2.5% of the boundary were categorized to belong to the guard band. Assume the cost of testing a MEMS accelerometer for all the specifications at each temperature is one dollar. If only the devices in the guard-band region are fully tested, then after elimination of the temperature tests, the total cost is only expected to be $(916 + 84×3) or $1168 as compared to the original test cost of $(1000 + 774×2) or $2548, a savings of almost 54%.

## 6. Conclusions

We have proposed a statistical learning based specification compaction procedure. As opposed to *ad hoc* compaction in use today, this methodology allows for a controlled yield loss and defect escape. Additionally, use of simulation for compaction is advantageous in terms of cost. We have also proposed the use of classification based $\varepsilon$-SVM modeling that is specifically suited for test. Initial application of the approach to an operational amplifier and a MEMS accelerometer shows significant test compaction can be achieved with very low yield loss and defect escape due to prediction error.

In future work, we plan to generate training instances that model the manufacturing process in a more accurate fashion and test instances that also contain real defects. We also intend to estimate the guard-band region based on the device distribution as opposed to a fixed value. Finally, we would like to build a cost model that accurately quantifies the reduction in test cost as well as overall device cost due to test compaction.

## Acknowledgement

The authors would like to thank Prof. Gary Fedder and Dr. Nilmoni Deb for their insights in temperature variation modeling and temperature test for the MEMS accelerometer.